\documentclass[a4paper,11pt]{article}

\makeatletter
\gdef\@fpheader{   }
\gdef\@journal{jhep}

\RequirePackage{amsmath}
\RequirePackage{amssymb}
\RequirePackage{epsfig}
\RequirePackage{CJK}
\RequirePackage{graphicx}
\RequirePackage[numbers,sort&compress]{natbib}
\RequirePackage{color}
\RequirePackage[colorlinks=true
,urlcolor=blue
,anchorcolor=blue
,citecolor=blue
,filecolor=blue
,linkcolor=blue
,menucolor=blue
,pagecolor=blue
,linktocpage=true
,pdfproducer=medialab
,pdfa=true
,CJKbookmarks
]{hyperref}

\newif\ifnotoc\notocfalse
\newif\ifemailadd\emailaddfalse
\newif\iftoccontinuous\toccontinuousfalse

\def\@subheader{\@empty}
\def\@keywords{\@empty}
\def\@abstract{\@empty}
\def\@xtum{\@empty}
\def\@dedicated{\@empty}
\def\@arxivnumber{\@empty}
\def\@collaboration{\@empty}
\def\@collaborationImg{\@empty}
\def\@proceeding{\@empty}
\def\@preprint{\@empty}

\newcommand{\subheader}[1]{\gdef\@subheader{#1}}
\newcommand{\keywords}[1]{\if!\@keywords!\gdef\@keywords{#1}\else%
\PackageWarningNoLine{\jname}{Keywords already defined.\MessageBreak Ignoring last definition.}\fi}
\renewcommand{\abstract}[1]{\gdef\@abstract{#1}}
\newcommand{\dedicated}[1]{\gdef\@dedicated{#1}}
\newcommand{\arxivnumber}[1]{\gdef\@arxivnumber{#1}}
\newcommand{\proceeding}[1]{\gdef\@proceeding{#1}}
\newcommand{\xtumfont}[1]{\textsc{#1}}
\newcommand{\correctionref}[3]{\gdef\@xtum{\xtumfont{#1} \href{#2}{#3}}}
\newcommand\jname{JHEP}
\newcommand\acknowledgments{\section*{Acknowledgments}}

\newcommand\preprint[1]{\gdef\@preprint{\hfill #1}}

\newcommand\note[2][]{%
\if!#1!%
\stepcounter{footnote}\footnotetext{#2}%
\else%
{\renewcommand\thefootnote{#1}%
\footnotetext{#2}}%
\fi}

\newtoks\auth@toks
\renewcommand{\author}[2][]{%
  \if!#1!%
    \auth@toks=\expandafter{\the\auth@toks#2\ }%
  \else
    \auth@toks=\expandafter{\the\auth@toks#2$^{#1}$\ }%
  \fi
}

\newtoks\affil@toks\newif\ifaffil\affilfalse
\newcommand{\affiliation}[2][]{%
\affiltrue
  \if!#1!%
    \affil@toks=\expandafter{\the\affil@toks{\item[]#2}}%
  \else
    \affil@toks=\expandafter{\the\affil@toks{\item[$^{#1}$]#2}}%
  \fi
}

\newtoks\email@toks\newcounter{email@counter}%
\newcommand{\emailAdd}[1]{%
\emailaddtrue%
\ifnum\theemail@counter>0\email@toks=\expandafter{\the\email@toks, \@email{#1}}%
\else\email@toks=\expandafter{\the\email@toks\@email{#1}}%
\fi\stepcounter{email@counter}}
\newcommand{\@email}[1]{\href{mailto:#1}{\tt #1}}

\newcommand*\collaboration[1]{\gdef\@collaboration{#1}}
\newcommand*\collaborationImg[2][]{\gdef\@collaborationImg{#2}}

\newcommand\afterLogoSpace{\smallskip}
\newcommand\afterSubheaderSpace{\vskip3pt plus 2pt minus 1pt}
\newcommand\afterProceedingsSpace{\vskip21pt plus0.4fil minus15pt}
\newcommand\afterTitleSpace{\vskip23pt plus0.06fil minus13pt}
\newcommand\afterRuleSpace{\vskip23pt plus0.06fil minus13pt}
\newcommand\afterCollaborationSpace{\vskip3pt plus 2pt minus 1pt}
\newcommand\afterCollaborationImgSpace{\vskip3pt plus 2pt minus 1pt}
\newcommand\afterAuthorSpace{\vskip5pt plus4pt minus4pt}
\newcommand\afterAffiliationSpace{\vskip3pt plus3pt}
\newcommand\afterEmailSpace{\vskip16pt plus9pt minus10pt\filbreak}
\newcommand\afterXtumSpace{\par\bigskip}
\newcommand\afterAbstractSpace{\vskip16pt plus9pt minus13pt}
\newcommand\afterKeywordsSpace{\vskip16pt plus9pt minus13pt}
\newcommand\afterArxivSpace{\vskip3pt plus0.01fil minus10pt}
\newcommand\afterDedicatedSpace{\vskip0pt plus0.01fil}
\newcommand\afterTocSpace{\bigskip\medskip}
\newcommand\afterTocRuleSpace{\bigskip\bigskip}
\newlength{\affiliationsSep}
\newcommand\beforetochook{\pagestyle{myplain}\pagenumbering{roman}}

\DeclareFixedFont\trfont{OT1}{phv}{b}{sc}{11}

\renewcommand\maketitle{
\pagestyle{empty}
\thispagestyle{titlepage}
\setcounter{page}{0}
\noindent{\small\scshape\@fpheader}\@preprint\par
\afterLogoSpace
\if!\@subheader!\else\noindent{\trfont{\@subheader}}\fi
\afterSubheaderSpace
\if!\@proceeding!\else\noindent{\sc\@proceeding}\fi
\afterProceedingsSpace
{\LARGE\flushleft\sffamily\bfseries\@title\par}
\afterTitleSpace
\hrule height 1.5\p@%
\afterRuleSpace
\if!\@collaboration!\else
{\Large\bfseries\sffamily\raggedright\@collaboration}\par
\afterCollaborationSpace
\fi
\if!\@collaborationImg!\else
{\normalsize\bfseries\sffamily\raggedright\@collaborationImg}\par
\afterCollaborationImgSpace
\fi
{\bfseries\raggedright\sffamily\the\auth@toks\par}
\afterAuthorSpace
\ifaffil\begin{list}{}{%
\setlength{\leftmargin}{0.28cm}%
\setlength{\labelsep}{0pt}%
\setlength{\itemsep}{\affiliationsSep}%
\setlength{\topsep}{-\parskip}}
\itshape\small%
\the\affil@toks
\end{list}\fi
\afterAffiliationSpace
\ifemailadd %% if emailadd is true
\noindent\hspace{0.28cm}\begin{minipage}[l]{.9\textwidth}
\begin{flushleft}
\textit{E-mail:} \the\email@toks
\end{flushleft}
\end{minipage}
\else %% if emailaddfalse do nothing
\PackageWarningNoLine{\jname}{E-mails are missing.\MessageBreak Plese use \protect\emailAdd\space macro to provide e-mails.}
\fi
\afterEmailSpace
\if!\@xtum!\else\noindent{\@xtum}\afterXtumSpace\fi
\if!\@abstract!\else\noindent{\renewcommand\baselinestretch{.9}\textsc{Abstract:}}\ \@abstract\afterAbstractSpace\fi
\if!\@keywords!\else\noindent{\textsc{Keywords:}} \@keywords\afterKeywordsSpace\fi
\if!\@arxivnumber!\else\noindent{\textsc{ArXiv ePrint:}} \href{http://arxiv.org/abs/\@arxivnumber}{\@arxivnumber}\afterArxivSpace\fi
\if!\@dedicated!\else\vbox{\small\it\raggedleft\@dedicated}\afterDedicatedSpace\fi
\ifnotoc\else
\iftoccontinuous\else\newpage\fi
\beforetochook\hrule
\tableofcontents
\afterTocSpace
\hrule
\afterTocRuleSpace
\fi
\setcounter{footnote}{0}
\pagestyle{myplain}\pagenumbering{arabic}
} % close the \renewcommand\maketitle{

\renewcommand{\baselinestretch}{1.1}\normalsize
\divide\@tempdima\baselineskip
\@tempcnta=\@tempdima
\skip\@mpfootins = \skip\footins
\renewcommand{\@dotsep}{10000}

\newcommand\ps@myplain{
\pagenumbering{arabic}
\renewcommand\@oddfoot{\hfill-- \thepage\ --\hfill}
\renewcommand\@oddhead{}}
\let\ps@plain=\ps@myplain

\newcommand\ps@titlepage{\renewcommand\@oddfoot{}\renewcommand\@oddhead{}}

\numberwithin{equation}{section}

\renewcommand\section{\@startsection{section}{1}{\z@}%
                                   {-3.5ex \@plus -1.3ex \@minus -.7ex}%
                                   {2.3ex \@plus.4ex \@minus .4ex}%
                                   {\normalfont\large\bfseries}}
\renewcommand\subsection{\@startsection{subsection}{2}{\z@}%
                                   {-2.3ex\@plus -1ex \@minus -.5ex}%
                                   {1.2ex \@plus .3ex \@minus .3ex}%
                                   {\normalfont\normalsize\bfseries}}
\renewcommand\subsubsection{\@startsection{subsubsection}{3}{\z@}%
                                   {-2.3ex\@plus -1ex \@minus -.5ex}%
                                   {1ex \@plus .2ex \@minus .2ex}%
                                   {\normalfont\normalsize\bfseries}}
\renewcommand\paragraph{\@startsection{paragraph}{4}{\z@}%
                                   {1.75ex \@plus1ex \@minus.2ex}%
                                   {-1em}%
                                   {\normalfont\normalsize\bfseries}}
\renewcommand\subparagraph{\@startsection{subparagraph}{5}{\parindent}%
                                   {1.75ex \@plus1ex \@minus .2ex}%
                                   {-1em}%
                                   {\normalfont\normalsize\bfseries}}

\def\fnum@figure{\textbf{\figurename\nobreakspace\thefigure}}
\def\fnum@table{\textbf{\tablename\nobreakspace\thetable}}

\long\def\@makecaption#1#2{%
  \vskip\abovecaptionskip
  \sbox\@tempboxa{\small #1. #2}%
  \ifdim \wd\@tempboxa >\hsize
    \small #1. #2\par
  \else
    \global \@minipagefalse
    \hb@xt@\hsize{\hfil\box\@tempboxa\hfil}%
  \fi
  \vskip\belowcaptionskip}

\renewenvironment{thebibliography}[1]{%
\begin{oldthebibliography}{#1}%
\small%
\raggedright%
\setlength{\itemsep}{5pt plus 0.2ex minus 0.05ex}%
}%
{%
\end{oldthebibliography}%
}

\makeatother

\usepackage[T1]{fontenc} % if needed
\usepackage{romannum} 
\usepackage{CJK}

\title{{\boldmath Duality family of KdV equation}}
\author[a]{Xin Gu,}
\author[b]{Yuan-Yuan Liu,}
\author[c,1]{Wen-Du Li,}\note{liwendu@tjnu.edu.cn}
\author[a,2]{and Wu-Sheng Dai}\note{daiwusheng@tju.edu.cn}

\affiliation[a]{Department of Physics, Tianjin University, Tianjin 300350, P.R. China}
\affiliation[b]{Theoretical Physics Division, Chern Institute of Mathematics, Nankai University, PR China}

\affiliation[c]{College of Physics and Materials Science, Tianjin Normal University, Tianjin 300387, PR China}

\abstract{It is revealed that there exist duality families of the KdV type equation. A
duality family consists of an infinite number of generalized KdV (GKdV)
equations. A duality transformation relates the GKdV equations in a duality
family. Once a family member is solved, the duality transformation presents
the solutions of all other family members. We show some dualities as examples,
such as the soliton solution-soliton solution duality and the periodic
solution-soliton solution duality.

}

\begin{document} %正文开始
\begin{CJK*}{GBK}{song}
\maketitle %生成题目

\flushbottom
%(正文开始) ――――――――――――――――――――――――――――――――――――――――――――――――――

\section{Introduction}

After Russell found the solitary wave phenomenon, studying nonlinear evolution
equations began in physics and mathematics \cite{ablowitz1991solitons}. When
Kortoweg and de Vries studied the water wave in the long-wave approximation
and finite small amplitude, they gave the Korteweg-de Vries (KdV) equation
\cite{kordeweg1895change,ablowitz1991solitons,peregrine1966calculations},%
\begin{equation}
\frac{\partial u}{\partial t}-6u\frac{\partial u}{\partial x}+\frac
{\partial^{3}u}{\partial x^{3}}=0. \label{KdV}%
\end{equation}
The KdV equation is a basic model in nonlinear evolution equations
\cite{karakoc2021new,silem2021nonisospectral}. The KdV equation defines many
physical phenomena, such as waves in anharmonic crystals
\cite{zabusky1967synergetic}, waves in bubble liquid mixtures
\cite{van1968equations}, ion acoustic waves
\cite{konno1974modified,haas2003quantum,schamel1973modified}, and waves in
warm plasma \cite{konno1974modified,haas2003quantum,schamel1973modified}.

\textit{Soliton solution. }The solitary wave solutions of the KdV equation are
noted as solitons. The velocity of the solitary wave relates to its magnitude
\cite{faddeev1978quantum}, and after the collision, it retains the original
magnitude, shape, and velocity \cite{korkmaz2010numerical,lamb1980elements}.
The theory of solitons emerges in biochemistry, nonlinear optics, mathematical
biosciences, fluid dynamics, plasma physics, nuclear physics, and geophysics
\cite{biswas20081}. There have been many approaches to calculating the soliton
solution \cite{wang1996application,kudryashov1988exact}, such as the
Painlev\'{e} analysis method, the B\"{a}cklund transformation method, the
Hirota bilinear method, the inverse scattering method, and the Darboux
transformation method \cite{ablowitz1991solitons}. These methods apply not
only to calculating the soliton solution of the KdV equation but also to other
partial differential equations \cite{dorfman1993dirac}. These methods have
different limits in applications, and there is no universal method for solving
nonlinear partial differential equations generally \cite{drazin1989solitons}.

\textit{Modified KdV (mKdV) equation and generalized KdV (GKdV) equation. }The
KdV equation is a special case of the GKdV equation. The GKdV equation is
\cite{melo2010generalized}
\begin{equation}
\frac{\partial u}{\partial t}-f\left(  u\right)  \frac{\partial u}{\partial
x}+\frac{\partial^{3}u}{\partial x^{3}}=0. \label{GKdV1}%
\end{equation}
The GKdV equation recovers the KdV equation (\ref{KdV}) when $f\left(
u\right)  =6u$.

A special GKdV equation with $f\left(  u\right)  =-\alpha u^{k}$ is the KdV
type equation with a power-law nonlinearity \cite{wazwaz2008new},%
\begin{equation}
\frac{\partial u}{\partial t}+\alpha u^{k}\frac{\partial u}{\partial x}%
+\frac{\partial^{3}u}{\partial x^{3}}=0, \label{powerKdV}%
\end{equation}
and the mKdV equation is (\ref{powerKdV}) with $k=2$ and $\alpha=6$
\cite{zhang2014solutions}. The Miura transformation establishes a one-to-one
correspondence between the solutions of the KdV equation and the solutions of
the mKdV equation \cite{miura1968korteweg}. The mKdV equation has a rich
physical background \cite{zhang2019wronskian,zhao2019rational}. The mKdV
equation can describe\ a bounded particle propagating in a one-dimensional
nonlinear lattice with a harmonic force \cite{wadati1975wave}, small amplitude
ion acoustic waves propagating in plasma physics \cite{konno1974modified}, and
the thermal pulse propagating through a\ single crystal of sodium fluoride
\cite{narayanamurti1970nonlinear,tappert1970asymptotic}.

\textit{Duality and duality family. }Newton in \textit{Principia} revealed a
duality between gravitation and elasticity in classical mechanics, called the
Newton-Hooke duality \cite{chandrasekhar2003newton}. E. Kasner and V.I.
Arnol'd independently find the generalized duality between power potentials:
two power potentials $U\left(  r\right)  =\xi r^{a}$ and $V\left(  r\right)
=\eta r^{A}$ are dual\ if $\frac{a+2}{2}=\frac{2}{A+2}$, called the
Kasner-Arnol'd theorem
\cite{arnold1990huygens,needham1998visual,arnol2013mathematical}.

Recently, we find that such a duality generally exists in classical mechanics,
quantum mechanics, and scalar fields and present the duality among arbitrary
potentials \cite{li2021duality}. We find that the duality is not a duality
only between two potentials but exists duality families \cite{li2021duality}.
Each duality family consists of infinite potentials; in a duality family,
every potential is dual to all other potentials. Once a family member's
solution is obtained, we can obtain all other members' solutions by the
duality transformation. Therefore, the duality relation can be used to find
the solutions for classical mechanics, quantum mechanics, field theory, and
nonlinear equations (such as the Gross-Pitaevskii equation)
\cite{li2022solving,chen2022indirect,liu2021exactly}. The duality can also be
used to classify long-range potentials in quantum mechanics \cite{li2021long}.

In this paper, we reveal the duality and duality families of the GKdV
equation. The duality transformation can transform the solution of a GKdV
equation into the solution of its dual GKdV equation. The GKdV equation
duality family consists of an infinite number of GKdV equations that are dual
to each other. The solution of all GKdV equations in a duality family can be
obtained from the solution of one solved family member by the duality
transformation. This way, we can obtain a series of exact solutions of GKdV
equations. As an example, we discuss the KdV equation duality family in which
the KdV equation (\ref{KdV}) and the KdV type equation with a power-law
nonlinearity (\ref{powerKdV}) are family members. The duality transformation
gives a series of 1-soliton solutions of GKdV equations from a 1-soliton
solution of the KdV equation (\ref{KdV}). We also consider the duality between
the periodic solution of the KdV equation and the soliton solution\ of the
mKdV equation.

In particular, since the solution of all GKdV equations in a duality family
can be obtained from the solution of one family member by the duality
transformation, we can develop an indirect approach for solving GKdV
equations: (1) constructing the duality family of a GKdV equation; (2) looking
for an `easy' equation in the duality family and solving the `easy' equation;
(3) solving the wanted equation by the duality transformation.

In section \ref{Section2}, we\ present the duality\ and duality family of the
GKdV equation. In section \ref{Section3}, we consider two examples: (1)
solving the KdV equation with a power-law nonlinearity from the KdV equation
by the duality transformation; (2) the duality between the periodic solution
of the KdV equation and the soliton solution\ of the mKdV equation. The
conclusion is given in section \ref{Section4}. In Appendix, we solve a
periodic solution of the KdV equation.

\section{Duality family of GKdV equation \label{Section2}}

In this section, we give the duality\ and duality family of the traveling wave
GKdV equation. The solutions of a GKdV equation can be obtained from its dual
equation by the duality transformation.

The traveling wave with a velocity $C$\ of the GKdV equation (\ref{GKdV1}) is
given by
\begin{equation}
\frac{d^{3}u}{dz^{3}}+\left[  C-f\left(  u\right)  \right]  \frac{du}{dz}=0.
\label{GKdV}%
\end{equation}
where $u\left(  x,t\right)  =u\left(  z\right)  $ and $z=x+Ct$.

The traveling wave GKdV equation (\ref{GKdV}) has the following duality relation.

\textit{Two traveling wave GKdV equations,}%
\begin{align}
\frac{d^{3}u}{dz^{3}}+\left[  C-f\left(  u\right)  \right]  \frac{du}{dz}  &
=0,\label{GKDV2}\\
\frac{d^{3}v}{d\zeta^{3}}+\left[  \mathcal{C}-g\left(  v\right)  \right]
\frac{dv}{d\zeta}  &  =0, \label{GKDV3}%
\end{align}
\textit{if}%
\begin{equation}
\frac{1}{C}u^{-2}\left[  G-U\left(  u\right)  -Fu\right]  =\frac
{1}{\mathcal{C}}v^{-2}\left[  \mathcal{G}-\mathcal{V}\left(  v\right)
-\mathcal{F}v\right]  , \label{PotenDual}%
\end{equation}
\textit{where}%
\begin{align}
\frac{d^{2}U\left(  u\right)  }{du^{2}}  &  =-f\left(  u\right)  ,\text{
\ }\label{dUg}\\
\frac{d^{2}\mathcal{V}\left(  v\right)  }{dv^{2}}  &  =-g\left(  v\right)  ,
\label{dUg2}%
\end{align}%
\begin{align}
F  &  =-\left[  \frac{d^{2}u}{dz^{2}}+Cu+\frac{dU\left(  u\right)  }%
{du}\right]  ,\label{DefF1}\\
\mathcal{F}  &  =-\left[  \frac{d^{2}v}{d\zeta^{2}}+\mathcal{C}v+\frac
{d\mathcal{V}\left(  v\right)  }{dv}\right]  ,
\end{align}
and%
\begin{align}
G  &  =\frac{1}{2}\left(  \frac{du}{dz}\right)  ^{2}+\frac{1}{2}%
Cu^{2}+U\left(  u\right)  +Fu,\label{DefG1}\\
\mathcal{G}  &  =\frac{1}{2}\left(  \frac{dv}{d\zeta}\right)  ^{2}+\frac{1}%
{2}\mathcal{C}v^{2}+\mathcal{V}\left(  v\right)  +\mathcal{F}v,
\end{align}
\textit{then their solutions satisfy}%
\begin{align}
u  &  \leftrightarrow v^{\sigma},\label{SolveDual1}\\
z  &  \leftrightarrow\sqrt{\frac{\mathcal{C}}{C}}\sigma\zeta.
\label{SolveDual2}%
\end{align}
\textit{Here }$\sigma$\textit{ is an arbitrarily chosen constant.}

\textit{Integral of motion. }Before going on, we first illustrate the meaning
of $G$, $F$, $\mathcal{G}$, and $\mathcal{F}$, taking $G$ and $F$ as examples.

Broadly speaking, $G$ and $F$\ are both integrals of motion for the equation
of motion\ (\ref{GKDV2}). In principle, the integral of the equation of motion
over time is known as the integral of motion. Here $G$ and $F$ are integration
constants of integrating the traveling wave equation (\ref{GKDV2}) over\ $z$
and $u$, respectively; we here still call them integral of motion.

Multiplying both sides of the GKdV equation (\ref{GKDV2}) by $dz$ and
integrating and using (\ref{dUg}) give $\frac{d^{2}u}{dz^{2}}+Cu+\frac
{dU\left(  u\right)  }{du}=-F$, i.e., (\ref{DefF1}), where $F$ is the
integration constant of the integral over $z$.

Similarly, multiplying both sides of (\ref{DefF1}) by $du$ and integrating
give $\frac{1}{2}\left(  \frac{du}{dz}\right)  ^{2}+\frac{1}{2}Cu^{2}+U\left(
u\right)  +Fu=G$, i.e., (\ref{DefG1}), where $G$ is the integration constant
of the integral over $u$ and $\int du\frac{d^{2}u}{dz^{2}}=\int dz\frac
{du}{dz}\frac{d^{2}u}{dz^{2}}=\frac{1}{2}\int dz\frac{d}{dz}\left(  \frac
{du}{dz}\right)  ^{2}=\frac{1}{2}\left(  \frac{du}{dz}\right)  ^{2}$ is used.

\textit{Proof of duality relation. }Substituting the duality transformations
(\ref{SolveDual1}) and (\ref{SolveDual2}) into (\ref{DefF1}) gives%
\begin{equation}
\frac{C}{\mathcal{C}}\frac{d^{2}v}{d\zeta^{2}}+\frac{C}{\mathcal{C}}\left(
\sigma-1\right)  v^{-1}\left(  \frac{dv}{d\zeta}\right)  ^{2}+\sigma
Cv+v^{2\left(  1-\sigma\right)  }\frac{dU\left(  v^{\sigma}\right)  }%
{dv}+\sigma v^{1-\sigma}F=0. \label{0.2}%
\end{equation}
By (\ref{DefG1}), we have
\begin{equation}
\frac{C}{\mathcal{C}}\left(  \sigma-1\right)  v^{-1}\left(  \frac{dv}{d\zeta
}\right)  ^{2}=2\left(  \sigma-1\right)  v^{1-2\sigma}\left[  G-U\left(
v^{\sigma}\right)  -Fv^{\sigma}\right]  -C\left(  \sigma-1\right)  v.
\label{0.2a}%
\end{equation}
Using (\ref{0.2a}) to eliminate the term $\left(  \sigma-1\right)
v^{-1}\left(  \frac{dv}{d\zeta}\right)  ^{2}$ in (\ref{0.2}), we arrive at%
\begin{equation}
\frac{C}{\mathcal{C}}\frac{d^{2}v}{d\zeta^{2}}+Cv+2\left(  \sigma-1\right)
v^{1-2\sigma}\left[  G-U\left(  v^{\sigma}\right)  -Fv^{\sigma}\right]
+v^{2\left(  1-\sigma\right)  }\frac{dU\left(  v^{\sigma}\right)  }{dv}+\sigma
v^{1-\sigma}F=0. \label{0.1}%
\end{equation}
By the duality transformation (\ref{PotenDual}), we can obtain%
\begin{equation}
\mathcal{V}\left(  v\right)  =\mathcal{G}-\mathcal{F}v-\frac{\mathcal{C}}%
{C}v^{2-2\sigma}\left[  G-U\left(  v^{\sigma}\right)  -Fv^{\sigma}\right]  .
\label{VDual}%
\end{equation}
Taking the derivative of (\ref{VDual})\ with respect to $v$ gives%
\begin{equation}
\frac{d\mathcal{V}\left(  v\right)  }{dv}=-\mathcal{F}+2\frac{\mathcal{C}}%
{C}\left(  \sigma-1\right)  v^{1-2\sigma}\left[  G-U\left(  v^{\sigma}\right)
-Fv^{\sigma}\right]  +\frac{\mathcal{C}}{C}v^{2\left(  1-\sigma\right)
}\left[  \frac{dU\left(  v^{\sigma}\right)  }{dv}+\sigma v^{\sigma-1}F\right]
. \label{V1}%
\end{equation}
Substituting (\ref{V1}) into (\ref{0.1}) gives%
\begin{equation}
\frac{d^{2}v}{d\zeta^{2}}+\mathcal{C}v+\frac{d\mathcal{V}\left(  v\right)
}{dv}+\mathcal{F}=0.
\end{equation}
Then taking the derivative with respect to $\zeta$ and using (\ref{dUg2}), we
arrive at (\ref{GKDV3}).

\textit{Discussion of }$U$. The relation between $f\left(  u\right)  $\ in the
GKdV equation (\ref{GKDV2}) and $U\left(  u\right)  $\ in (\ref{dUg}) is not
unique. $U\left(  u;a,b\right)  =U\left(  u\right)  +au+b$ and $U\left(
u\right)  $ lead to the same $f\left(  u\right)  $, and both correspond to the
GKdV equation (\ref{GKdV1}).

The integral of motion $F$, corresponding to $U\left(  u;a,b\right)  $, by
(\ref{DefF1}), is $F\left(  a,b\right)  =-\left[  \frac{d^{2}u}{dz^{2}%
}+Cu+\frac{dU\left(  u;a,b\right)  }{du}\right]  =F-a$; the integral of motion
$G$, corresponding to $U\left(  u;a,b\right)  $ , by (\ref{DefG1}), is
$G\left(  a,b\right)  =\frac{1}{2}\left(  \frac{du}{dz}\right)  ^{2}+\frac
{1}{2}Cu^{2}+U\left(  u;a,b\right)  +F\left(  a,b\right)  u=G+b$. Therefore,
by (\ref{PotenDual}), the duality transformation\ given by $U\left(
u;a,b\right)  $\ is%
\begin{equation}
\frac{1}{C}u^{-2}\left[  G\left(  a,b\right)  -U\left(  u;a,b\right)
-F\left(  a,b\right)  u\right]  =\frac{1}{\mathcal{C}}v^{-2}\left[
\mathcal{G}-\mathcal{V}\left(  v;a,b\right)  -\mathcal{F}v\right]  .
\label{PotenDual2'}%
\end{equation}
Here $\mathcal{V}\left(  v;a,b\right)  $ is the duality\ of $U\left(
u;a,b\right)  $.

Substituting $U\left(  u;a,b\right)  $, $F\left(  a,b\right)  $, and $G\left(
a,b\right)  $ into the duality transformation (\ref{PotenDual2'}) gives%
\begin{equation}
\mathcal{V}\left(  v;a,b\right)  =\mathcal{G}-\mathcal{F}v-\frac{\mathcal{C}%
}{C}v^{2-2\sigma}\left[  G-U\left(  v^{\sigma}\right)  -Fv^{\sigma}\right]
=\mathcal{V}\left(  v\right)  .
\end{equation}
That is, in the GKdV equation, although the correspondence between $f\left(
u\right)  $ and $U\left(  u\right)  $ is not unique, the same $f\left(
u\right)  $ corresponding to different\ $U\left(  u\right)  $, the choice\ of
$U\left(  u\right)  $ does not influence the duality of the GKdV equation.

\section{Duality family of KdV equation: Example \label{Section3}}

We consider a special duality family of the GKdV equation as an example. The
KdV equation and mKdV equation are family members of this duality family. The
solutions of all family members in a duality family are related by a duality
transformation. In a duality family containing the KdV equation, we can solve
all the GKdV equations in the family from the solution of the KdV equation by
the duality transformation. In this section, we give the solution of the KdV
equation with a power-law nonlinearity from the solution of the KdV equation;
the mKdV equation is the power-law nonlinearity KdV equation with power $2$.

\textit{Duality family of the KdV equation and the KdV equation with a
power-law nonlinearity. }The KdV equation (\ref{KdV}) with $z=x-Ct$,%
\begin{equation}
\frac{d^{3}u}{dz^{3}}-\left(  C+6u\right)  \frac{du}{dz}=0, \label{KdV1}%
\end{equation}
has a $1$-soliton solution \cite{griffiths2010traveling}%
\begin{equation}
u\left(  z\right)  =-\frac{C}{2}\operatorname{sech}^{2}\left(  \frac{\sqrt{C}%
}{2}z\right)  . \label{1-solitonWave}%
\end{equation}
The soliton solution\ is a localized traveling wave solution. The localization
solution, taking the $1$-soliton solution as an example, means that
(\ref{1-solitonWave}) when $z\rightarrow\pm\infty$, $u\left(  z\right)
\rightarrow0$. The integral of motion of\ the $1$-soliton solution
(\ref{1-solitonWave}), by (\ref{DefF1}), (\ref{DefG1}) and
(\ref{1-solitonWave}), is%
\begin{equation}
F=0\text{ \ and\ \ }G=0. \label{FG}%
\end{equation}
Then the dual equation of the traveling wave KdV equation\ given by the
duality transformation (\ref{PotenDual}) is%
\begin{equation}
\frac{d^{3}v}{d\zeta^{3}}-\left[  \mathcal{C}+\frac{\mathcal{C}}{C}\left(
2+\sigma\right)  \left(  1+\sigma\right)  v^{\sigma}\right]  \frac{dv}{d\zeta
}=0. \label{FamilyOfKdV}%
\end{equation}
Since $\sigma$ can be chosen arbitrarily, (\ref{FamilyOfKdV}) is not a single
equation but forms a duality family. All the GKdV equations labeled by
different $\sigma$ in the duality family are dual equations of the KdV equation.

By (\ref{SolveDual1})\ and (\ref{SolveDual2}), we can obtain the solution of
(\ref{FamilyOfKdV})
\begin{equation}
v\left(  \zeta\right)  =\left[  -\frac{C}{2}\operatorname{sech}^{2}\left(
\frac{\sqrt{\mathcal{C}}}{2}\sigma\zeta\right)  \right]  ^{1/\sigma},
\label{sGKdV}%
\end{equation}
where $\zeta=x-\mathcal{C}t$ has a velocity $-\mathcal{C}$.

Instead of $z$, represent the dual equation (\ref{FamilyOfKdV}) by $\left(
t,x\right)  $:
\begin{equation}
\frac{\partial v}{\partial t}+\alpha v^{\sigma}\frac{\partial v}{\partial
x}+\frac{\partial^{3}v}{\partial x^{3}}=0, \label{FamilyOfKdVP}%
\end{equation}
where $\alpha=-\frac{\mathcal{C}}{C}\left(  2+\sigma\right)  \left(
1+\sigma\right)  $. When $\sigma$ is taken as a positive integer,
(\ref{FamilyOfKdVP}) is the KdV equation with a power-law nonlinearity, and
the solution (\ref{sGKdV})\ becomes%
\begin{equation}
v\left(  x,t\right)  =\left\{  -\frac{C}{2}\operatorname{sech}^{2}\left[
\frac{\sqrt{\mathcal{C}}}{2}\sigma\left(  x-\mathcal{C}t\right)  \right]
\right\}  ^{1/\sigma}, \label{FamilySoluXingBo}%
\end{equation}
or equivalently, $v\left(  x,t\right)  =\left\{  \frac{\mathcal{C}\left(
2+\sigma\right)  \left(  1+\sigma\right)  }{2\alpha\cosh^{2}\left[
\frac{\sqrt{\mathcal{C}}}{2}\sigma\left(  x-\mathcal{C}t\right)  \right]
}\right\}  ^{1/\sigma}$, which agrees with Ref. \cite{hayek2010constructing}.

In this duality family, the family member $\sigma=1$ is the KdV equation
(\ref{KdV}), and the family member $\sigma=2$ is the mKdV equation
\begin{equation}
\frac{\partial v}{\partial t}-12\frac{\mathcal{C}}{C}v^{2}\frac{\partial
v}{\partial x}+\frac{\partial^{3}v}{\partial x^{3}}=0. \label{mKdV}%
\end{equation}
(\ref{FamilySoluXingBo}) with $\sigma=2$\ gives the $1$-soliton solution of
the mKdV equation (\ref{mKdV})
\begin{equation}
v\left(  x,t\right)  =\pm\sqrt{-\frac{C}{2}}\operatorname{sech}\left[
\sqrt{\mathcal{C}}\left(  x-\mathcal{C}t\right)  \right]  .
\end{equation}
Now, by the duality relation, we have obtained all family members' solutions
from the KdV equation's solution.

\textit{Periodic solution-soliton solution duality. }A duality exists between
the periodic solution and the soliton solution of the GKdV equation. We take
the periodic solution of the KdV equation and the soliton solution\ of the
mKdV equation as an\ example.

The KdV equation (\ref{KdV}) has a periodic solution
\begin{equation}
u\left(  x,t\right)  =\frac{1}{6}C\left\{  1+3\tan^{2}\left[  \frac{\sqrt{C}%
}{2}\left(  x-Ct\right)  \right]  \right\}  . \label{ZhouQiJie}%
\end{equation}
The KdV equation (\ref{KdV}) with $z=x-Ct$ becomes (\ref{KdV1}), and its
solution (\ref{ZhouQiJie}) becomes
\begin{equation}
u\left(  z\right)  =\frac{C}{6}\left[  1+3\tan^{2}\left(  \frac{C}{2}z\right)
\right]
\end{equation}
with the period $\frac{2\pi}{\sqrt{C}}$.

The integral of motion of the periodic solution (\ref{ZhouQiJie}) of the KdV
equation, by (\ref{DefF1}), (\ref{DefG1}) and (\ref{ZhouQiJie}), is
\begin{equation}
F=0,\text{ \ }G=-\frac{C^{3}}{54}.
\end{equation}
The dual equation of the traveling wave KdV equation given by the duality
transformation (\ref{PotenDual}) is then
\begin{equation}
\frac{d^{3}v}{d\zeta^{3}}+\left[  \mathcal{C}-\frac{1}{27}\left(
1-\sigma\right)  \left(  1-2\sigma\right)  \mathcal{C}C^{2}v^{-2\sigma}%
+\frac{\mathcal{C}}{C}\left(  \sigma+1\right)  \left(  \sigma+2\right)
v^{\sigma}\right]  \frac{dv}{d\zeta}=0, \label{FamilyOfKdVZhouQi}%
\end{equation}
where $\zeta=x+\mathcal{C}t$. The duality transformations (\ref{SolveDual1})
and (\ref{SolveDual2}) give the solution of (\ref{FamilyOfKdVZhouQi}).%
\begin{equation}
v\left(  \zeta\right)  =\left\{  \frac{C}{6}\left[  1-3\tanh^{2}\left(
\frac{\sqrt{\mathcal{C}}}{2}\sigma\zeta\right)  \right]  \right\}  ^{1/\sigma
}. \label{FamilySoluXingBoZhouqi}%
\end{equation}
$\sigma$ running over all possible values gives all equations and their
solutions in the duality family.

The family member $\sigma=1$ and $\mathcal{C}=-C$ in the duality family is the
KdV equation (\ref{KdV}). Different from the $1$-soliton solution
(\ref{FamilyOfKdV}), however, the family member $\sigma=-1$ is the traveling
wave mKdV equation
\begin{equation}
\frac{d^{3}v}{d\zeta^{3}}+\mathcal{C}\left(  1-\frac{2}{9}C^{2}v^{2}\right)
\frac{dv}{d\zeta}=0. \label{XingboMKdV}%
\end{equation}
or, with $\zeta=x+\mathcal{C}t$ and $\mathcal{C}=\frac{27}{C^{2}}$,
\begin{equation}
\frac{\partial v}{\partial t}-6v^{2}\frac{\partial v}{\partial x}%
+\frac{\partial^{3}v}{\partial x^{3}}=0, \label{0.4}%
\end{equation}
which, by (\ref{FamilySoluXingBoZhouqi}), has a traveling wave solution
\begin{equation}
v\left(  x,t\right)  =\frac{2\sqrt{\mathcal{C}}}{\sqrt{3}\left\{  1-3\tanh
^{2}\left[  \frac{\sqrt{\mathcal{C}}}{2}\left(  x+\mathcal{C}t\right)
\right]  \right\}  }. \label{0.3}%
\end{equation}
It can be directly verified that $v\left(  x,t\right)  \rightarrow-\frac
{\sqrt{3\mathcal{C}}}{3}$ when $x,t\rightarrow\pm\infty$, so (\ref{0.3})\ is a
soliton solution\ of the mKdV equation (\ref{0.4}).

In this example, the duality of the periodic solution is a soliton solution.

\textit{Indirect approach for solving equations. }The above example inspires
us to develop an indirect approach to solving equations. When solving an
equation, we can (1) find its duality family; (2) look for and solve an `easy'
family member, and (3) achieve the solution of this equation by the duality transformation.

\section{Conclusion \label{Section4}}

This paper reveals a duality among the GKdV equations, and all the GKdV
equations that are dual to each other form a duality family. In a duality
family, the solutions of different family members are related by the duality transformation.

In a duality family, we only need to solve one family member, and the duality
transformation can give solutions for all other family members. This allows us
to develop an indirect approach to solving the GKdV equation.

In this paper, as an example, we discuss the GKdV equation duality family
containing the KdV equation and the KdV equation with a power-law
nonlinearity: seeking $1$-soliton solution of the KdV equation with a
power-law nonlinearity from a $1$-soliton solution of the KdV equation by the
duality relation. In another example, we consider the periodic
solution-soliton solution duality. By the duality transformation, we give a
soliton solution\ of the mKdV equation from a periodic solution of the KdV equation.

\appendix

\section{Appendix Periodic solution of KdV equation}

The KdV equation%
\begin{equation}
\frac{\partial u}{\partial t}-6u\frac{\partial u}{\partial x}+\frac
{\partial^{3}u}{\partial x^{3}}=0,
\end{equation}
with $z=x-Ct$ converts into%
\begin{equation}
\frac{d^{3}u}{dz^{3}}-\left(  C+6u\right)  \frac{du}{dz}=0.
\end{equation}

Multiplying both sides by $dz$ and integrating give%
\begin{equation}
\frac{d^{2}u}{dz^{2}}-Cu-3u^{2}=-F.
\end{equation}
Then multiplying by $du$ and integrating give%
\begin{equation}
\frac{1}{2}\left(  \frac{du}{dz}\right)  ^{2}-\frac{1}{2}Cu^{2}-u^{3}+Fu=G,
\label{0.1}%
\end{equation}
where $\int du\frac{d^{2}u}{dz^{2}}=\frac{1}{2}\left(  \frac{du}{dz}\right)
^{2}$ is used.

Let $x=u\left(  z\right)  $ and $y=\frac{du\left(  z\right)  }{dz}$, and then
(\ref{0.1}) is converted into an equation of a cubic algebraic curve
\begin{equation}
y^{2}=2x^{3}+Cx^{2}-2Fx+2G. \label{0.2}%
\end{equation}
Taking the transformation%
\begin{align}
x^{\prime}  &  =x+\frac{1}{6}C,\nonumber\\
y^{\prime}  &  =\sqrt{2}y
\end{align}
converts (\ref{0.2}) into an elliptic curve in Weierstrass normal form
\begin{equation}
y^{\prime2}=4x^{\prime3}-g_{2}x^{\prime}-g_{3}%
\end{equation}
with
\begin{align}
g_{2}  &  =\frac{C^{2}}{3}+4F,\nonumber\\
g_{3}  &  =-4G-\frac{2CF}{3}-\frac{C^{3}}{27}.
\end{align}

By the differential equation of the Weierstrass-$\wp$ function,
\begin{equation}
\left(  \wp^{\prime}\right)  ^{2}=4\wp^{3}-g_{2}\wp-g_{3},
\end{equation}
we can give the solution of the differential equation (\ref{0.1})
\begin{equation}
u\left(  z\right)  =\wp\left(  \sqrt{\frac{1}{2}}\left(  z+z_{0}\right)
;\frac{C^{2}}{3}+4F,-4G-\frac{2CF}{3}-\frac{C^{3}}{27}\right)  -\frac{1}{6}C,
\end{equation}
denoted by the Weierstrass-$\wp$ function.

By relation
\begin{equation}
a^{2}\wp\left(  az;g_{2},g_{3}\right)  =\wp\left(  z;a^{4}g_{2},a^{6}%
g_{3}\right)  , \label{beishugongshi}%
\end{equation}
we have%
\begin{equation}
u\left(  z\right)  =2\wp\left(  z+z_{0};\frac{C^{2}}{12}+F,-\frac
{18CF+C^{3}+108G}{216}\right)  -\frac{1}{6}C.
\end{equation}
That is, the KdV equation has a traveling wave solution represented by the
Weierstrass-$\wp$ function
\begin{equation}
u\left(  x,t\right)  =2\wp\left(  x-Ct+\varphi_{0};\frac{C^{2}}{12}%
+F,-\frac{18CF+C^{3}+108G}{216}\right)  -\frac{1}{6}C, \label{0.3}%
\end{equation}
where $\varphi_{0}=z_{0}$\ is an initial phase.

When $g_{2}$\ and $g_{3}$ in $\left(  \wp^{\prime}\right)  ^{2}=4\wp^{3}%
-g_{2}\wp-g_{3}$ satisfy
\begin{equation}
g_{2}^{3}-27g_{3}^{2}=0,
\end{equation}
the Weierstrass-$\wp$ function\ reduces to a trigonometric or a hyperbolic function.

For the traveling wave solution (\ref{0.3}), $g_{2}^{3}-27g_{3}^{2}=0$\ gives%

\begin{equation}
-C^{2}F^{2}-16F^{3}+2C^{3}G+36CFG+108G^{2}=0. \label{0.4}%
\end{equation}
For simplicity, we take the integral of motion $F=0$, then Eq. (\ref{0.4})
becomes
\begin{equation}
C^{3}G+54G^{2}=0.
\end{equation}
That is, when the integral of motion $G=0$ or $G=-\frac{C^{2}}{54}$, the
traveling wave solution (\ref{0.3})\ reduces to a hyperbolic or a
trigonometric function.

When $G=0$ and $F=0$, the traveling wave solution (\ref{0.3}) becomes
\begin{equation}
u\left(  x,t\right)  =2\wp\left(  x-Ct+\varphi_{0};\frac{C^{2}}{12}%
,-\frac{C^{3}}{216}\right)  -\frac{1}{6}C.
\end{equation}
Taking $\varphi_{0}=i\pi$ gives%
\begin{equation}
u\left(  x,t\right)  =-\frac{1}{2}C\operatorname{sech}^{2}\left[  \frac
{\sqrt{C}}{2}\left(  x-Ct\right)  \right]  ,
\end{equation}
or, equivalently,%
\begin{equation}
u\left(  z\right)  =-\frac{1}{2}C\operatorname{sech}^{2}\left(  \frac{\sqrt
{C}}{2}z\right)  .
\end{equation}

When $G=-\frac{C^{2}}{54}$ and $F=0$, the traveling wave solution (\ref{0.3})
becomes%
\begin{equation}
u\left(  x,t\right)  =2\wp\left(  x-Ct+\varphi_{0};\frac{C^{2}}{12}%
,\frac{C^{3}}{216}\right)  -\frac{1}{6}C.
\end{equation}
Taking $\varphi_{0}=\pi$ gives%
\begin{equation}
u\left(  x,t\right)  =\frac{1}{6}C\left\{  1+3\tan^{2}\left[  \frac{\sqrt{C}%
}{2}\left(  x-Ct\right)  \right]  \right\}  ,
\end{equation}
or, equivalently,%
\begin{equation}
u\left(  z\right)  =\frac{1}{6}C\left[  1+3\tan^{2}\left(  \frac{\sqrt{C}}%
{2}z\right)  \right]  .
\end{equation}

Moreover, it is worthy to note that the elliptic curve is doubly-periodic
function. The KdV equation may have a doubly-periodic solution.

\acknowledgments

We are very indebted to Dr G. Zeitrauman for his encouragement. This work is supported in part by Special Funds for theoretical physics Research Program of the NSFC under Grant No. 11947124, and NSFC under Grant Nos. 11575125 and 11675119.

%(正文结束)――――――――――――――――――――――――――――――――――――――――――――――――――

%%%%%%%%%%%%%%%%%%%参考文献%%%%%%%%%%%%%%%%%%%%%%%%%%%

\providecommand{\href}[2]{#2}\begingroup\raggedright\endgroup

%%%%%%%%%%%%%%%%%%%bibtex形式的参考文献%%%%%%%%%%%%%%%
%\bibliographystyle{JHEP} %参考文献的风格(.bst)
%\bibliography{refs} %参考文献文件(.bib)
%\nocite{*} %若去掉注释，没有被引用的文献也被列出

%%%%%%%%%%%%%%%%%%%bbl形式的参考文献%%%%%%%%%%%%%%%%%%

%%%%%%%%%%%%%%%%%%%%%%%%%%%%%%%%%%%%%%%%%%%%%%%%%%%%%%

\end{CJK*}
\end{document}